\documentclass{PoS}

\title{Hadronic Resonance Production with ALICE Experiment at LHC}

\ShortTitle{Hadronic Resonance Production with ALICE Experiment at LHC}

\author{\speaker{S. Singha (for ALICE collaboration)}\\
        National Institute of Science Education and Research (NISER), Bhubaneswar\\
        E-mail: \email{subhash.singha@cern.ch}}

\abstract{The production of resonances in heavy-ion collisions is expected to be sensitive to the properties of strongly interacting matter  created in such collisions. We report on the measurements of $\phi$ and $K^{*0}$ resonances in Pb-Pb collisions at  $\sqrt{s_{NN}} $ = 2.76 TeV. The masses, widths and yields in Pb-Pb collisions as a function of centrality are compared to that in pp collisions to understand the role of re-scattering and regeneration. The resonance to non-resonance particle ratios are shown as a function of collision centrality and compared with the results at lower energies.}

\FullConference{8th International Workshop on Critical Point and Onset of Deconfinement,\\
		March 11 to 15, 2013\\
		Napa, California, USA}

\begin{document}

\section{Introduction}

The production of resonances plays an important role in understanding high energy heavy-ion collisions as their lifetime is short, comparable to the lifetime of the fireball created in these collisions  \cite{qgp}.  So their characteristic properties such as mass, width and yield as a function of collision centrality and transverse momentum ($p_{T}$) could be modified. This is possible because decay products of resonances may undergo re-scattering, resulting in a loss of the signal and on the other hand resonance regeneration may happen inside the medium. These two competing processes play an essential role in deciding the final resonance yields and resonance to non-resonance particle ratios \cite{partratio}. Here we present the results on $\phi$ and $K^{*0}$ mesons which probe the hadronic fireball in a different way due to a factor of 10 difference of their lifetime.

\section{Analysis details}
The analysis is done with the data in Pb-Pb collisions at $\sqrt{s_{NN}} $ = 2.76 TeV taken during 2010 with the ALICE \cite{alicedet} detector. The $\phi$ and $K^{*0}$ {\footnote[1]{The analysis is done for both $K^{*0}$ and anti-${K^{*0}}$}} are reconstructed via their hadronic decay channels ($\phi$ $\rightarrow$ $K^{+}$$K^{-}$ and $K^{*0}$ $\rightarrow$ $K^{\pm}\pi^{\mp}$) at mid-rapidity ($|y| < $ 0.5).  The pions and kaons are identified using the specific energy loss inside the volume of Time Projection Chamber (TPC).
The combinatorial background was evaluated using two different techniques, event-mixing technique and like-sign technique. The signal was obtained after the subtraction of combinatorial background. The signal is then fitted to a Breit-Wigner function for $K^{*0}$ or a Voigtian function for $\phi$, in addition a polynomial function is used to take care of residual backgrounds. The mass, width and yields are extracted from the fits for different $p_{T}$ bins in different collision centralities. The raw yields are corrected for detector acceptance, efficiency  and the branching ratios. The final corrected spectra are fitted with a Tsallis \cite{tsallis} function for the $K^{*0}$ {\footnote[2]{Since the $K^{*0}$ spectra is not well described by the Blast-Wave parameterization for the peripheral collisions.}} and a Boltzmann Gibbs Blast Wave \cite{bgbw} function for the $\phi$. The $p_{T}$ integrated invariant yields and the mean transverse momentum ($p_{T}$) were extracted from the data in the measured region and using the fit function for extrapolating  $p_{T}$ distributions to the remaining regions. 
\section{Results and Discussions}

The mass and width of $\phi$ and ${K^{*0}}$ extracted from the fit are shown in the Figure 1. Both masses and widths are consistent with the PDG values. The deviation of mass of ${K^{*0}}$ at low $p_{T}$,  also observed in pp collisions at $\sqrt{s}$ = 7 TeV, is a detector effect.

\begin{figure}[hbtp]
\begin{center}
\includegraphics[scale=0.09]{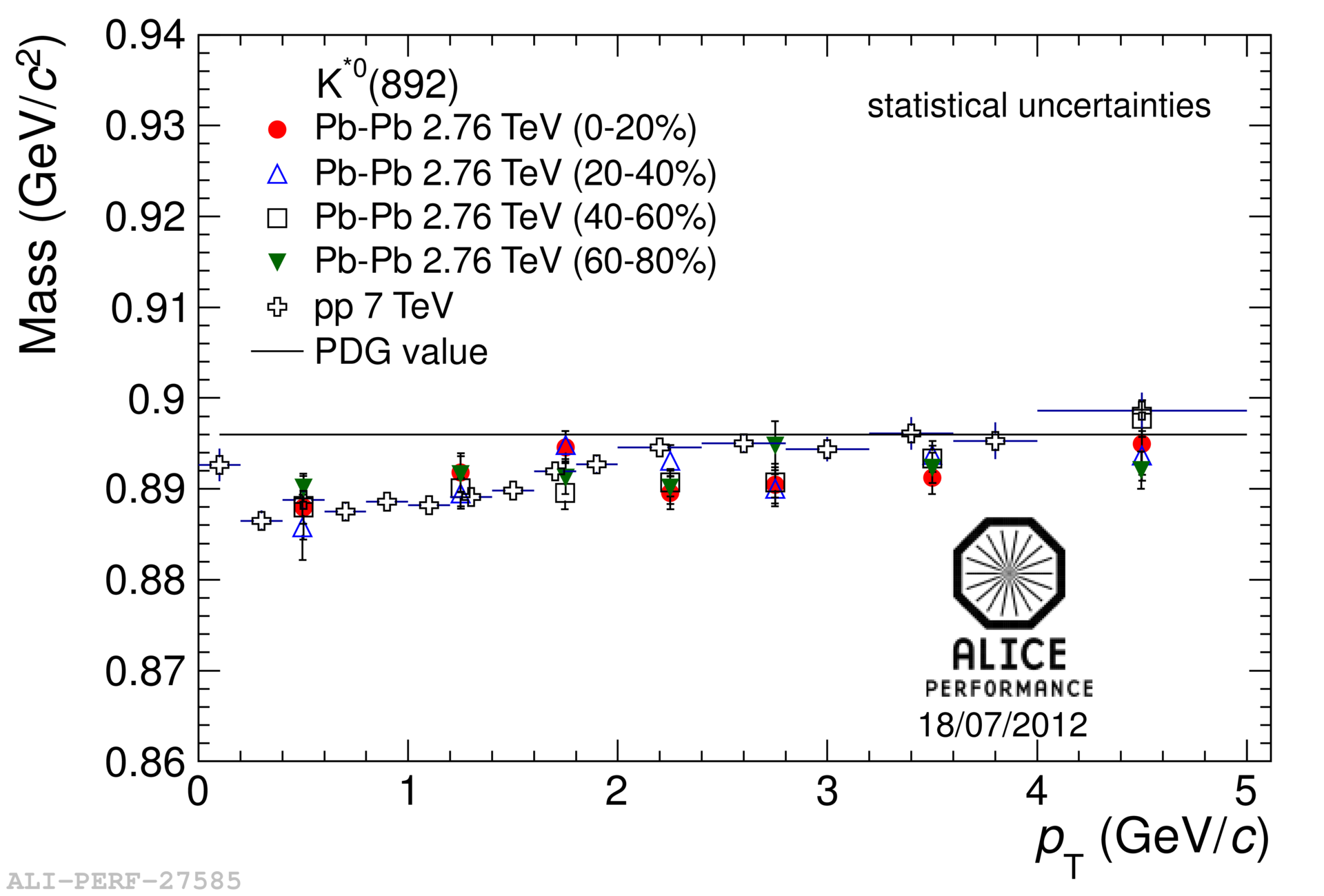}
\includegraphics[scale=0.09]{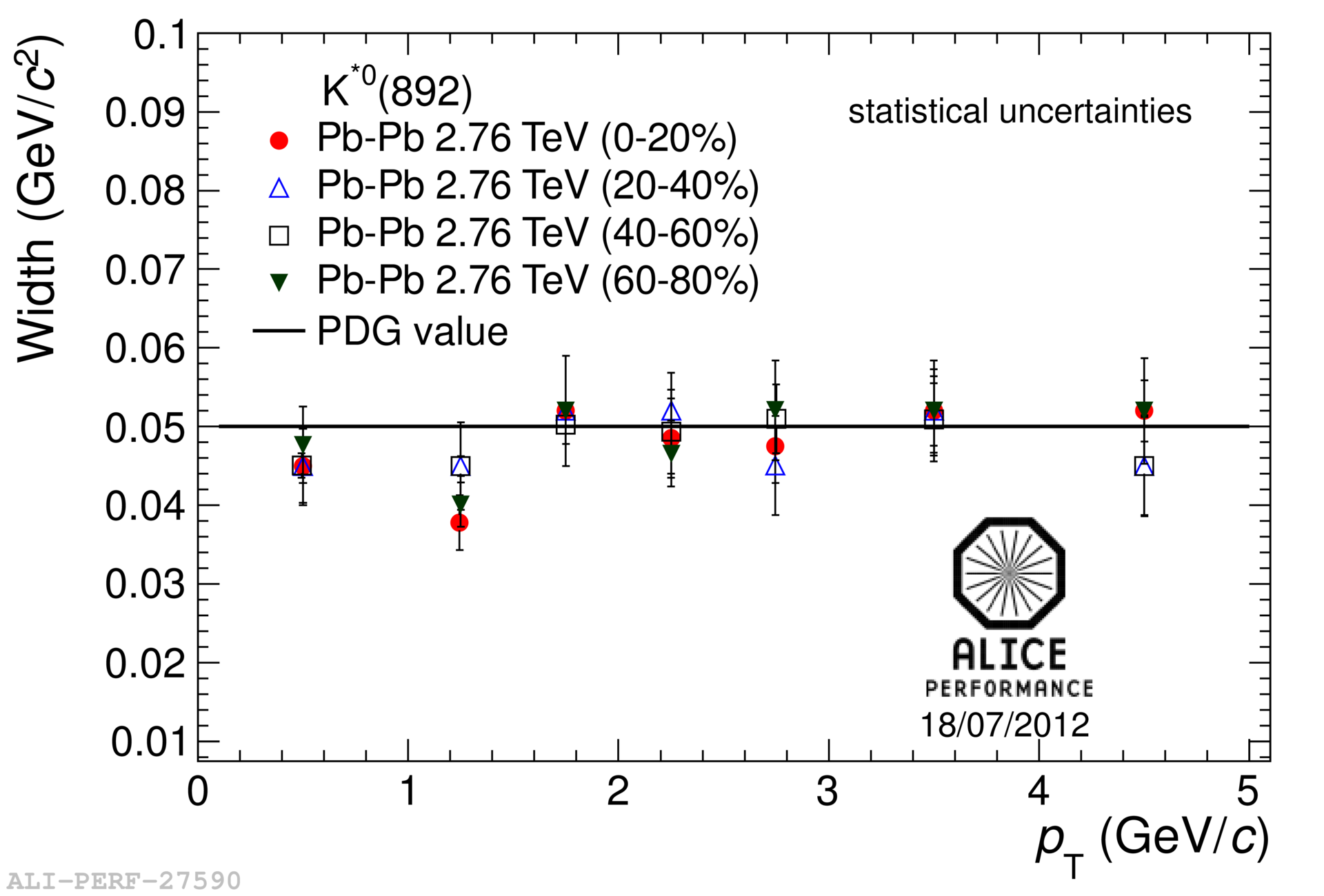}
\includegraphics[scale=0.09]{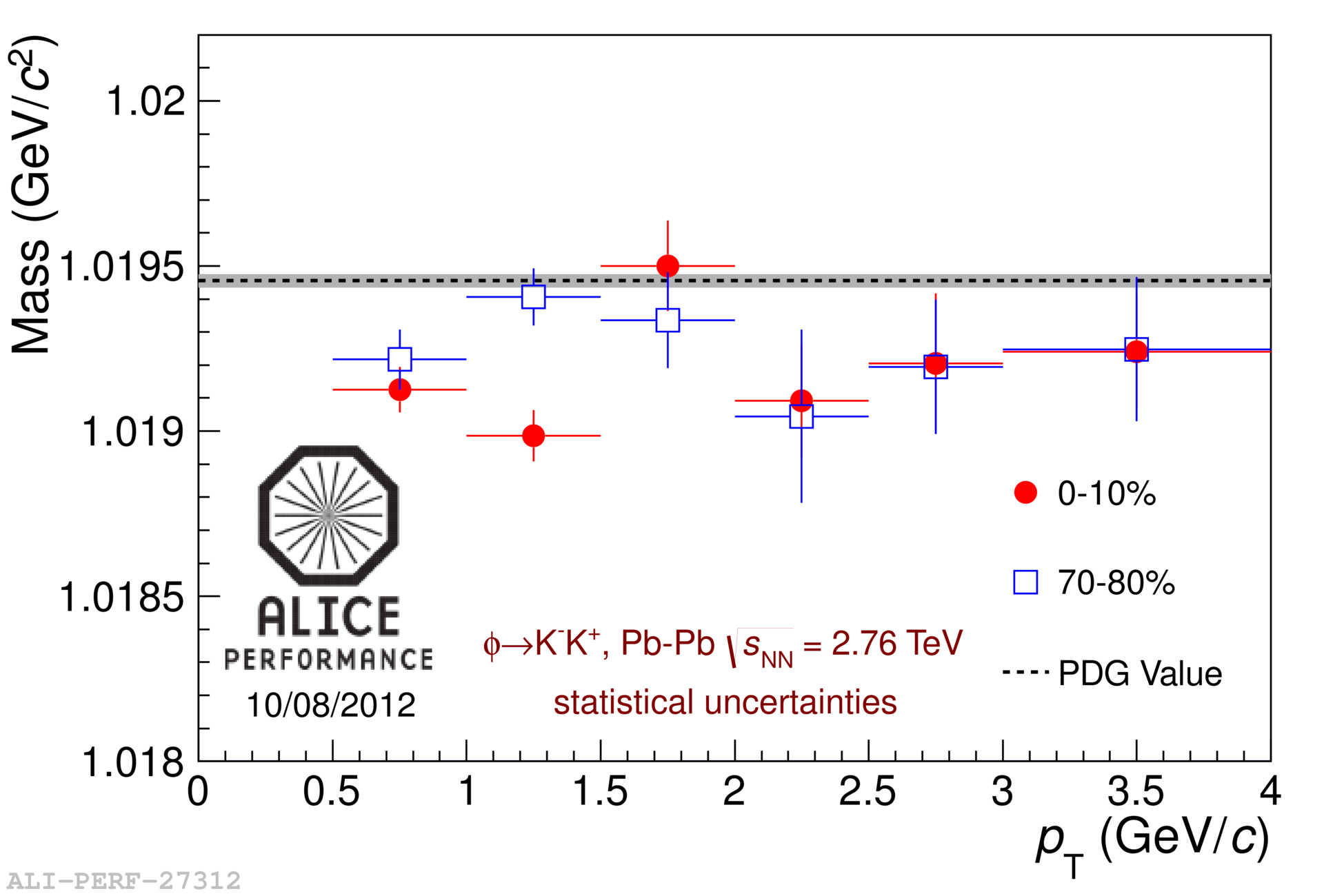}
\includegraphics[scale=0.09]{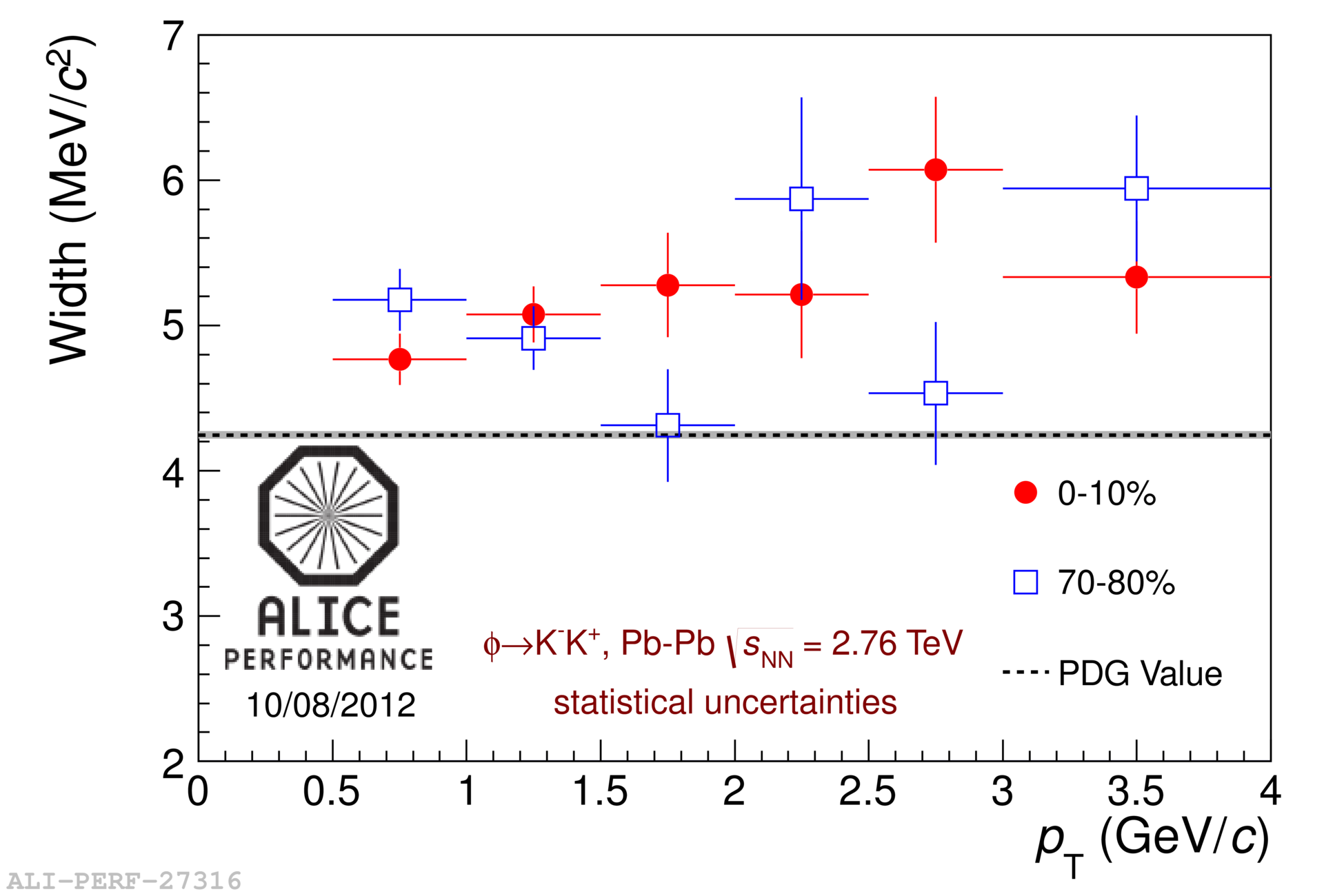}
\caption{(color online) Mass and width of  $\phi$ and $K^{*0}$ as a function of $p_{T}$ in Pb-Pb collisions at $\sqrt{s_{NN}}$ = 2.76 TeV.  The uncertainties shown are statistical only.}
\end{center}
\end{figure}

The centrality and energy dependence of particle ratios are studied for $K^{*0}/K^{-}$ and $\phi/K^{-}$ ratios. Figure 2 (left panel) shows the $K^{*0}/K^{-}$ and $\phi/K^{-}$ as a function of mean number of participating nucleons in Pb-Pb collisions. The $\phi/K^{-}$  ratio shows no dependence on collision centrality while a weak centrality dependence is observed in $K^{*0}/K^{-}$ ratio. This may be due to hadronic re-scattering for most central collisions, considering that the lifetime of $K^{*0}$ is factor of 10 smaller compared to that for $\phi$-meson. Figure 2 (right panel) shows that the $\phi/K$ ratio is independent of beam energy which may disfavor the production of $\phi$ mainly through kaon coalescence \cite{kaoncoal}.  
\begin{figure}[hbtp]
\begin{center}
\includegraphics[scale=0.09]{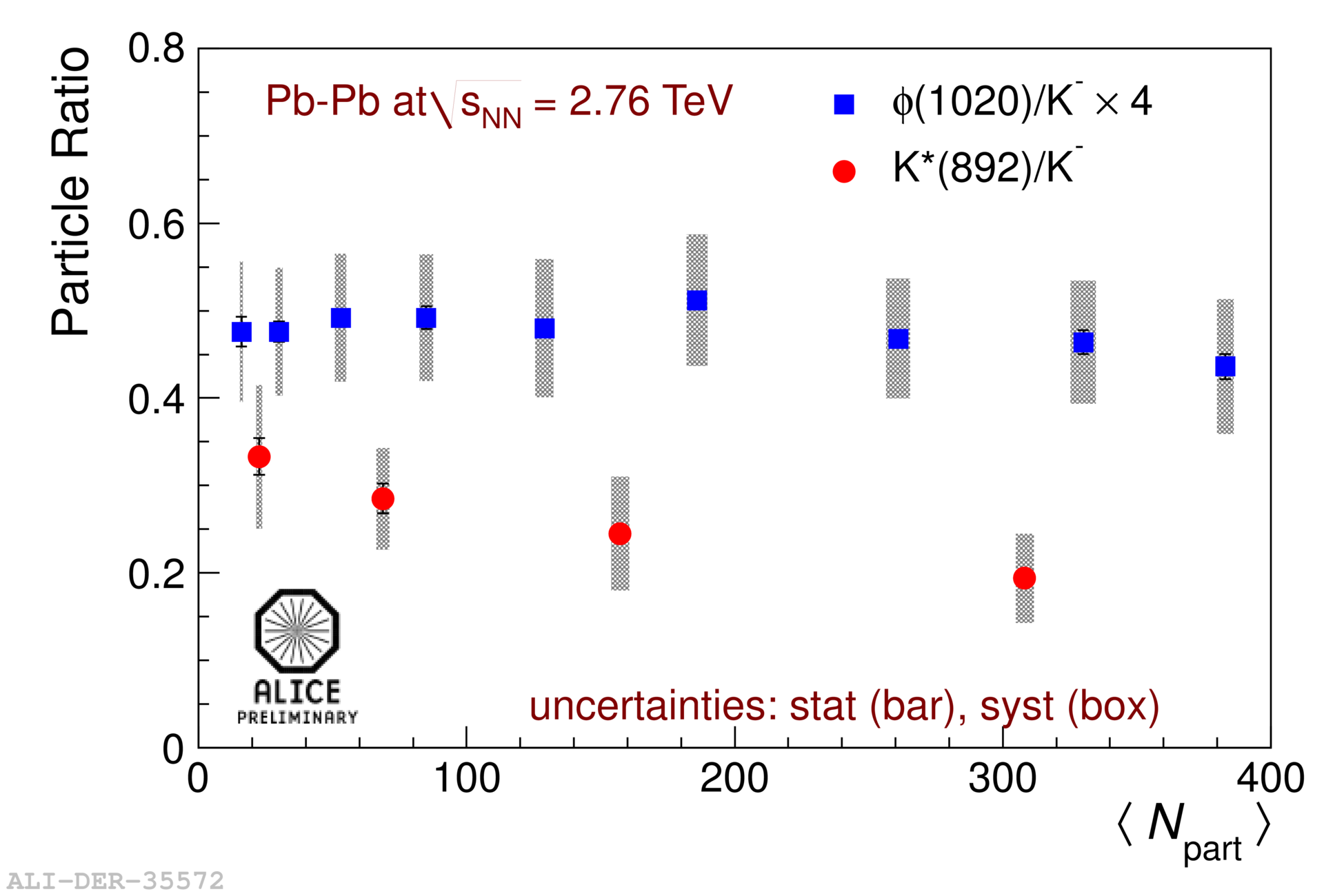}
\includegraphics[scale=0.09]{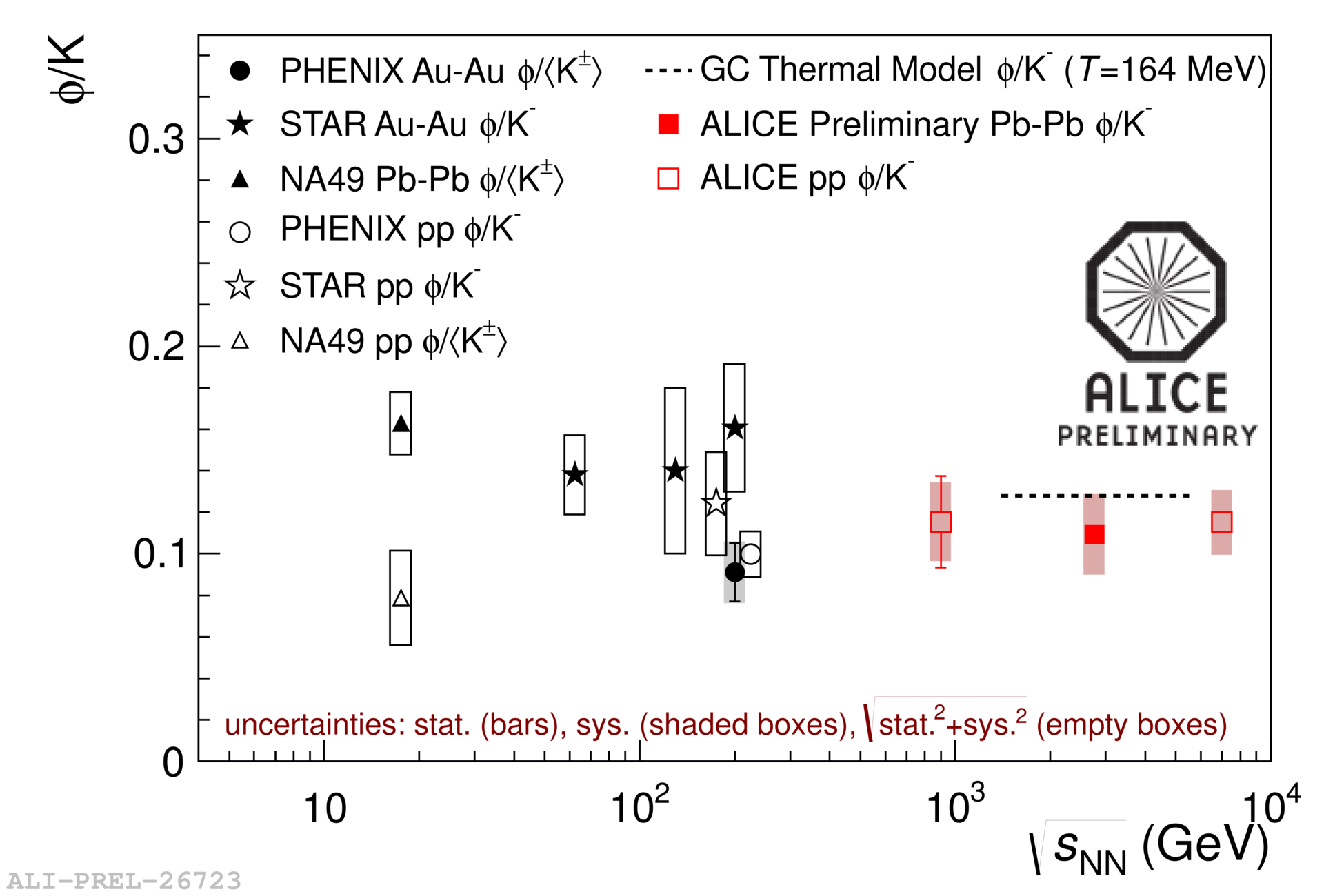}
\caption{(color online) Left panel: $\phi/K^{-}$ and $K^{*0}/K^{-}$ as a function of mean number of participating nucleons in Pb-Pb collisions at $\sqrt{s_{NN}}$ = 2.76 TeV.  The $K^{*0}/K^{-}$ points are shown by solid red circles while the $\phi/K^{-}$ points are shown by solid blue squares. Right panel: $\phi/K$ as a function of beam energy. The ALICE data (red symbols) \cite{ppalice} are compared with the measurements from SPS \cite{spspp} and RHIC \cite{rhicpp}. The statistical uncertainties are shown by bars while the systematic uncertainties are shown by shaded bands.}
\end{center}
\end{figure}

\begin{figure}[hbtp]
\begin{center}
\includegraphics[scale=0.09]{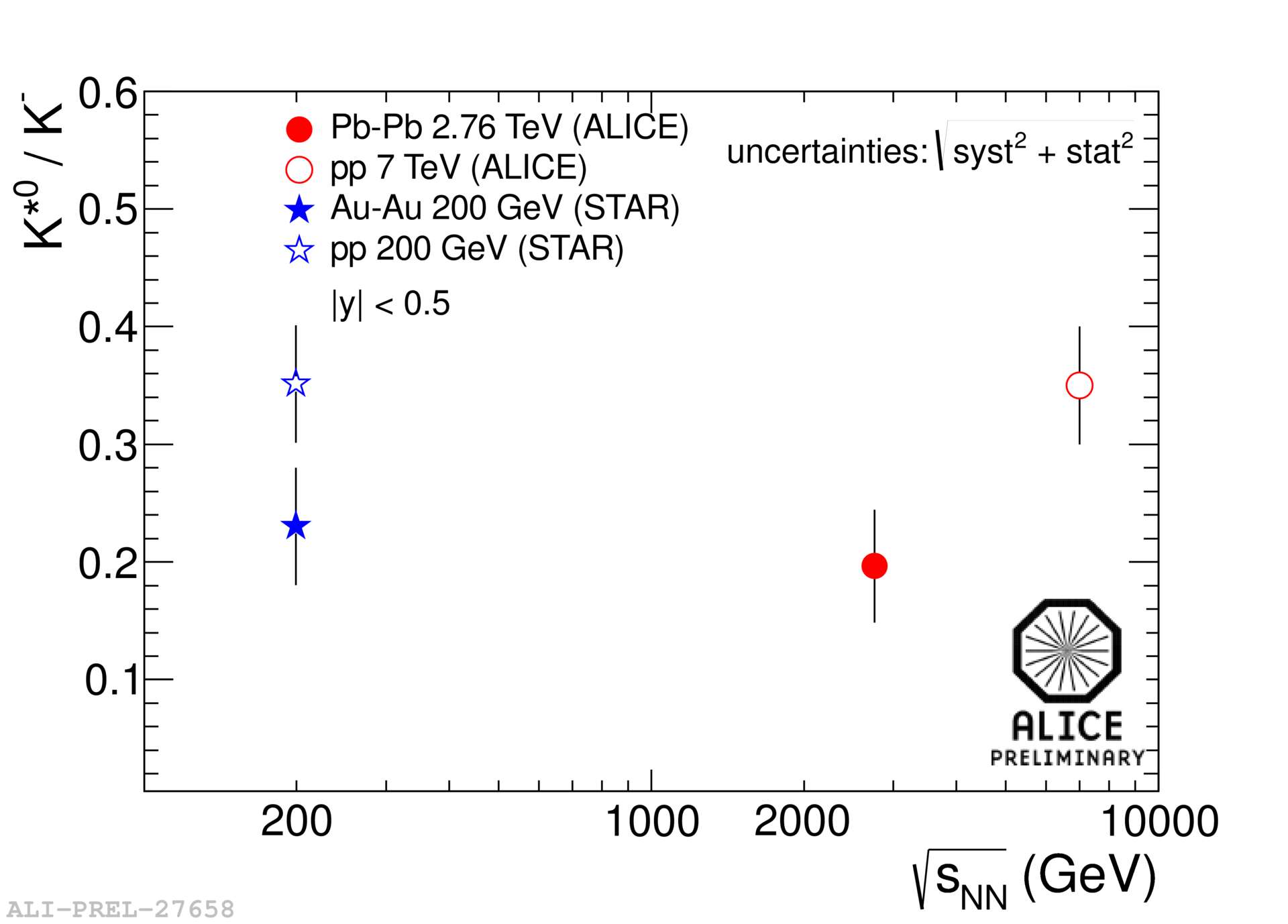}
\includegraphics[scale=0.095]{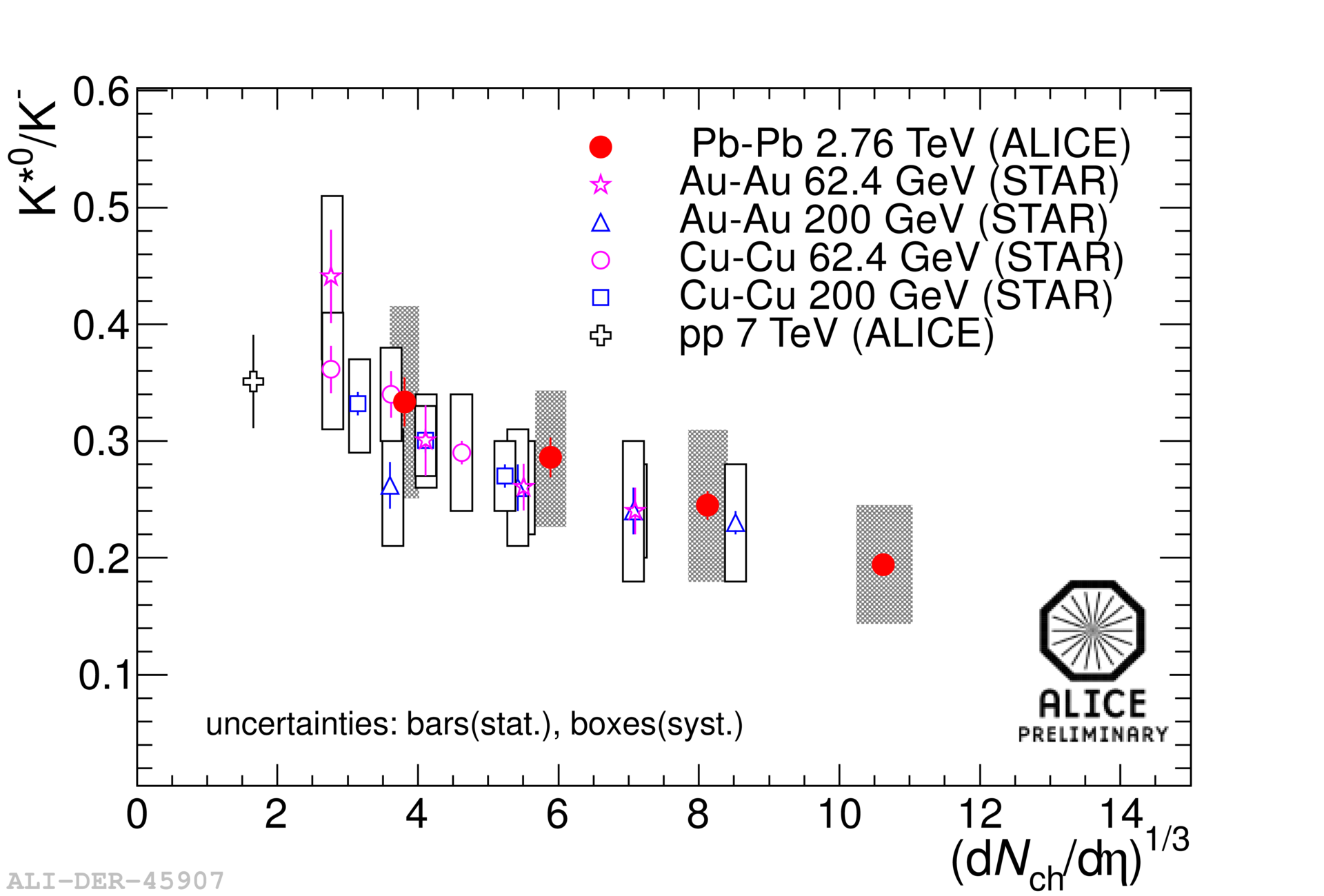}
\caption{(color online) Left panel: $K^{*0}/K^{-}$ as a function of beam energy. The ALICE data (red symbols)  \cite{ppalice} are compared with the measurements at RHIC \cite{rhicpp}. Right panel: $K^{*0}/K^{-}$ as a function of $(dN_{ch}/d\eta)^{1/3}$. The ALICE data (solid red symbols) at $\sqrt{s_{NN}}$ = 2.76 TeV in Pb-Pb collisions are compared with the STAR data at $\sqrt{s_{NN}}$ = 62.4 GeV and $\sqrt{s_{NN}}$ = 200 GeV in Cu-Cu and Au-Au collisions \cite{starphik}. The pp data at $\sqrt{s}$ = 7 TeV is shown by the open plus symbol.}
\end{center}
\end{figure}

Figure 3 (left panel) shows the $K^{*0}/K^{-}$ ratio in heavy-ion collisions is smaller than that measured in pp collisions (for both LHC and RHIC energies), indicating the presence of hadronic re-scattering in heavy-ion collisions. Figure 3 (right panel) shows the $K^{*0}/K^{-}$ ratio as a function of $(dN_{ch}/d\eta)^{1/3}$. The ALICE results are compared with the measurements at RHIC \cite{starphik}. The LHC data show the same trend as observed at RHIC. There is an indication of a further decrease at the highest $(dN_{ch}/d\eta)^{1/3}$ which may be attributed to a larger fireball size (and lifetime) accessible at the LHC energy.

\section{Summary}
In summary, we have presented a few selected results on $\phi$ and $K^{*0}$ meson production in Pb-Pb collisions at $\sqrt{s_{NN}}$ = 2.76 TeV. The masses and widths are found to be consistent with the PDG values. A weak centrality dependence observed in $K^{*0}/K^{-}$  while $\phi/K^{-}$ shows no centrality dependence. The decreasing trend in $K^{*0}/K^{-}$ may  point to the prevalence of the hadronic re-scattering over recombination for most central collisions. The $K^{*0}/K^{-}$ ratio in heavy-ion collisions is found to be lower than that measured in pp collisions, which is compatible with the influence of re-scattering. The $K^{*0}/K^{-}$ ratio studied as a function of $(dN_{ch}/d\eta)^{1/3}$ is found to decrease with increase in $(dN_{ch}/d\eta)^{1/3}$, which is equivalent to the previously shown centrality dependence.

\end{document}